\documentclass{aa}  

\usepackage[varg]{txfonts}
\usepackage{graphicx}
\usepackage{gensymb}
\usepackage{natbib,twoopt}
\usepackage[breaklinks=true]{hyperref} 
\usepackage{adjustbox}

\hypersetup{
           colorlinks=True,
           allcolors=blue
           }
\bibpunct{(}{)}{;}{a}{}{,} 
\makeatletter
\newcommandtwoopt{\citeads}[3][][]{\href{http://adsabs.harvard.edu/abs/#3}%
{\def\hyper@linkstart##1##2{}%
\let\hyper@linkend\@empty\citealp[#1][#2]{#3}}}
\newcommandtwoopt{\citepads}[3][][]{\href{http://adsabs.harvard.edu/abs/#3}%
{\def\hyper@linkstart##1##2{}%
\let\hyper@linkend\@empty\citep[#1][#2]{#3}}}
\newcommandtwoopt{\citetads}[3][][]{\href{http://adsabs.harvard.edu/abs/#3}%
{\def\hyper@linkstart##1##2{}%
\let\hyper@linkend\@empty\citet[#1][#2]{#3}}}
\newcommandtwoopt{\citeyearads}[3][][]%
{\href{http://adsabs.harvard.edu/abs/#3}
{\def\hyper@linkstart##1##2{}%
\let\hyper@linkend\@empty\citeyear[#1][#2]{#3}}}
\makeatother

\newcommand{\tu}{\textsuperscript}

\newcommand{\changeurlcolor}[1]{\hypersetup{urlcolor=#1}}
\usepackage{amstext}

\begin{document}

   \title{Location and energetics of the ultra-fast outflow \\ in PG 1448+273}
   
   \titlerunning{Location and energetics of the ultra-fast outflow in PG 1448+273}
   \authorrunning{M.\ Laurenti et al.}
    
   \subtitle{}

   \author{M.\ Laurenti\inst{1,2} \and A.\ Luminari\inst{1,2} \and F.\ Tombesi\inst{1,2,3,4} \and F.\ Vagnetti\inst{1,2} \and R.\ Middei\inst{2,5} \and E.\ Piconcelli\inst{2}}

   \institute{Dipartimento di Fisica, Università di Roma "Tor Vergata", Via della Ricerca Scientifica 1, 00133 Roma, Italy \\ \email{\changeurlcolor{black}\href{mailto:marco.laurenti@roma2.infn.it}{marco.laurenti@roma2.infn.it}} \and INAF - Osservatorio Astronomico di Roma, Via Frascati 33, 00040 Monte Porzio Catone, Italy \and NASA Goddard Space Flight Center, Greenbelt, MD 20771, USA \and Department of Astronomy, University of Maryland, College Park, MD 20742, USA \and Space Science Data Center, SSDC, ASI, Via del Politecnico snc, 00133 Roma, Italy} 

   \date{}

 
  \abstract
   {Ultra-fast outflows (UFOs) are the most powerful disk-driven winds in active galactic nuclei (AGNs). Theoretical and observational evidence shows that UFOs play a key role in the AGN feedback mechanism. The mechanical power of the strongest UFOs may be enough to propagate the feedback to the host galaxies and ultimately shape the AGN-galaxy coevolution. It is therefore of paramount importance to fully characterize UFOs, their location, and energetics.}  
   {We study two \emph{XMM-Newton} archival observations of the narrow-line Seyfert 1 galaxy (NLSy1) PG 1448+273. We concentrate on the latest observation, whose spectrum is characterized by a strong absorption feature in the Fe K band. This feature represents the spectral imprint of a UFO, as confirmed by other independent analyses. We study this feature in detail with a novel modeling tool. }
   {In order to constrain the physical properties of the UFO, we implemented the novel model called wind in the ionized nuclear environment (WINE) to fit the photoionized emission and absorption lines from a disk wind in X-ray spectra. WINE is a photoionization model that allows us to self-consistently calculate absorption and emission profiles. It also takes special relativistic effects into account.}
   {Our detection of the UFO in PG 1448+273 is very robust. The outflowing material is highly ionized, $\log\xi = 5.53_{-0.05}^{+0.04}$ erg s$^{-1}$ cm, has a high column density, $N_\mathrm{H} = 4.5_{-1.1}^{+0.8} \times 10^{23}$ cm$^{-2}$, is ejected with a maximum velocity $v_0 = 0.24^{+0.08}_{-0.06}\,c$ (90\% confidence level errors), and attains an average velocity $v_\mathrm{avg} = 0.152\,c$. WINE succeeds remarkably well to constrain a launching radius of $r_0=77_{-19}^{+31} \, r_\mathrm{S}$ from the black hole. We also derive a lower limit on both the opening angle of the wind ($\theta > 72\degree$) and the covering factor ($C_\mathrm{f} > 0.69$). 
   We find a mass outflow rate $\Dot{M}_\mathrm{out}=0.65^{+0.44}_{-0.33}\,M_\odot\,\mathrm{yr}^{-1} = 2.0^{+1.3}_{-1.0}\, \Dot{M}_\mathrm{acc}$ and a high instantaneous outflow kinetic power $\Dot{E}_\mathrm{out}=4.4^{+4.4}_{-3.6} \times 10^{44}$ erg s$^{-1}$ = 24\% $L_\mathrm{bol}$ = 18\% $L_\mathrm{Edd}$ ($1 \sigma$ errors). We find that a major error contribution on the energetics is due to $r_0$, stressing the importance of an accurate determination through proper spectral modeling, as done with WINE.
   Finally, using 20 \emph{Swift} (UVOT and XRT) observations together with the simultaneous Optical Monitor data from \emph{XMM-Newton}, we also find that $\alpha_\mathrm{ox}$ varied strongly, with a maximum excursion of $\Delta\alpha_\mathrm{ox} =-0.7$, after the UFO was detected, leading to a remarkable X-ray weakness. This may indicate a starving of the inner accretion disk due to the removal of matter through the wind, and it may have repercussions for the larger population of observed X-ray weak quasars.}
   {}


   \maketitle


\section{Introduction} 

Active galactic nuclei (AGNs) are known to host various outflow phenomena, many of which are observable in the X-ray band \citep[e.g.,][]{tombesi2013, kraemer2018, reeves2018, kallman2019, smith2019}. Some of them may reach very high velocities, up to a substantial fraction of the speed of light, and are thus called ultra-fast outflows \citep[UFOs;][]{tombesi2010a, chartas2014, nardini2015, tombesi2015, Matzeu2017, parker2017, Matzeu2019, reeves2020}. UFOs are typically detected thanks to blueshifted Fe K absorption lines, especially between 7 and 10 keV. These lines are produced by photoionized gas located in the circumnuclear regions of AGNs, and they are usually due to Fe XXV/XXVI K-shell resonant absorption.
UFOs have been detected in radio-quiet and radio-loud, and in both type I and II AGNs \citep[e.g.,][]{tombesi2010a, tombesi2010b, gofford2013, tombesi2014}. Because of current instrumental limitations, most of the UFOs have been found in low-$z$ objects, with some notable exceptions \citep[e.g.,][]{lanzuisi2012, vignali2015, chartas2016, dadina2018}.

Ultra-fast outflows are known to carry significant kinetic power, typically consisting of a few percent of the AGN bolometric luminosity, together with a mass flux of $\sim 0.01 - 1\, M_\odot\,\mathrm{yr}^{-1}$. Their energetics is thus high enough to affect the host galaxy \citep[e.g.,][]{hopkins2010, gaspari2011b, gaspari2011a}, the star formation, the supermassive black-hole (SMBH) growth, and the bulge evolution, offering an interpretation of the observed $M_\mathrm{BH}-\sigma$ relation \citep[e.g.,][]{ferrarese2000, dasyra2007}.

 \cite{kosec2020} recently reported evidence of a UFO in the X-ray spectrum of the narrow-line Seyfert 1 galaxy (NLSy1) PG 1448+273.
This source is located at redshift $z = 0.0645$, has a bolometric luminosity of $\log{L_\mathrm{bol}}=45.27$ erg s$^{-1}$ , and it is characterized by a considerable Eddington ratio of $\log\lambda_\mathrm{Edd} = -0.123$ (see Tab.\ \ref{tab:infoPG} for further information about PG 1448+273).
We unequivocally confirm this UFO based on an independent analysis, and we also constrain its physical parameters in a stringent way, including the launching radius and energetics. We achieve these results with the novel model called wind in the ionized nuclear environment (WINE), which is used here to simultaneously characterize the X-ray emission and absorption from the disk wind.

\noindent The standard $\Lambda$CDM cosmology ($H_\mathrm{0}=70$ km s$^{-1}$ Mpc$^{-1}$, $\Omega_\mathrm{m}=0.3$, $\Omega_\Lambda=0.7$) is adopted throughout the paper.

 {
    \begin{table}[t]
    \centering
    \caption{Overview of the observation.}
    \renewcommand{\arraystretch}{1.1}
    \begin{adjustbox}{max width=\columnwidth}
    \begin{tabular}{ c c c c}
    \hline\hline
     Obs.\ ID & Start date & Instrument & Net exp. \\
              & yyyy-mm-dd &            & (ks) \\
    \hline 
    \hline
    0781430101  &  2017-01-24  &   pn    &   76.3  \\
                &              & MOS1    &107.2\\
                &              &MOS2     &110.0\\
    \hline
    \end{tabular}
    \end{adjustbox}
    \label{tab:overview}
    \end{table}
    }

\section{Observations and data reduction} 
    
   We analyzed the most recent archival observation of the NLSy1 PG 1448+273 from the 
   \emph{XMM-Newton} satellite \citep{jansen2001} in detail. The observation was performed on January 24, 2017. \emph{XMM-Newton} also observed PG1448+273 on February 8, 2003. However, this older observation was characterized by a much shorter ($\sim 20$ ks) exposure, resulting in a lower quality dataset.
   In addition, while the spectrum related to the older observation is basically featureless, the spectrum associated with the latest observation is dominated by a clear, broad Fe K absorption feature attributable to a UFO.
   For this reason, because our primary goal is to model the UFO with the novel WINE model, our spectral analysis is focused on the more recent observation. The lack of significant UFO feature detection in the first \emph{XMM-Newton} observation was quantified in \citet{kosec2020}. Nevertheless, we also reduced this older observation in order to discuss its main properties in the context of spectral and $\alpha_\mathrm{ox}$ variability (see Sect.\ \ref{sec:aox}).      
    
    Raw data were retrieved from the \emph{XMM-Newton} Science Archive and then processed using the \emph{XMM-Newton} Science Analysis System (SAS v18.0.0) with the latest available calibration files. 
    We took advantage of the full potential of \emph{XMM-Newton} in the Fe K band by collecting data from all of its primary instruments, that is, the EPIC-pn and the two MOS cameras.
    After the canonical process of data reduction, which is fully described on the SAS web page, we binned the spectra in order to have at least 25 counts in each bin and a minimum group width of one-third of the instrumental energy resolution.
    This allowed us to adopt the $\chi^2$ statistics in a meaningful way.
    \noindent Tab.\ \ref{tab:overview} includes the overall properties of the observation we are interested in, and a general description of PG 1448+273 is given in Tab.\ \ref{tab:infoPG}.

    \begin{table*}[t]
        \centering
        \caption{Overall properties of PG 1448+273, including the SDSS identifier, the sky coordinates in degrees, the Galactic hydrogen column density, the values of bolometric luminosity, black-hole mass, and Eddington ratio from the catalogue of \citet{shen2011}, and the source redshift.}
        \renewcommand{\arraystretch}{1.5}
        \begin{tabular}{c c c c c c c c}
        \hline\hline
        SDSS ID & R.A. & Dec  & $N_\mathrm{H,gal}$ (10$^{20}$ cm$^{-2}$) & $\log{L_\mathrm{bol}}$ (erg s$^{-1}$) & $\log(M_\mathrm{BH}/M_\odot)$ &  $\log{\lambda_\mathrm{Edd}}$ &  $z$ \\
        \hline
        J145108.76+270926.9 & 222.787 & 27.157 & 3.01 & 45.27 & 7.29 &  $-0.123$ & 0.0645  \\
        \hline
        \end{tabular}
        \label{tab:infoPG}
    \end{table*}
    
    In addition, this analysis was complemented with data from the Optical Monitor \citep[OM;][]{mason2001}. 
    This instrument observed PG 1448+273 with the sole UVW2 (2120 \AA) filter in the latest observation.
    \noindent Raw OM data were converted into science products by invoking the SAS task \texttt{omichain}.
    In order to convert the spectral points into a suitable format for \texttt{XSPEC} \citep{arnaud1996}, we used the standard task \texttt{om2pha}.\\
    \indent The NLSy1 galaxy PG 1448+273 was also observed with the \emph{Neil Gehrels Swift Observatory} \citep[][ hereafter \emph{Swift}]{gehrels2004} for 20 epochs between 2010 and 2019 with both the XRT \citep{burrows2005} and UVOT \citep{roming2005} telescopes. These exposures provide a long-term simultaneous ultraviolet-X-ray monitoring, and raw data were retrieved from the multi-mission archive website\footnote{Available at \url{http://www.asdc.asi.it/mma.html}.}.
    Both X-ray (XRT) and optical-ultraviolet (UVOT) data were reduced using the standard tools provided by the Space Science Data
Center (SSDC) website. In particular, we performed an online interactive analysis for all the available observations with a sufficient exposure. XRT data were extracted using a circular region with radius $\sim$ 45 arcsec and an annulus centered on the source.  Spectra were grouped to have five counts per bin, and the Cash statistics \citep{Cash1979} was adopted for minimization purposes.
In addition, we used the UVOT aperture photometry to obtain the monochromatic fluxes for all the available filters. The source extraction region had a 5 arcsec radius, and a blank annular region centered on the source was used for the background.
 
\section{Spectral analysis and phenomenological modeling} 

    The \emph{XMM-Newton} spectral analysis of PG 1448+273 was carried out in two main steps.
    We first concentrated on the hard X-ray spectrum ($E = 2-9$ keV rest frame), in order to provide a detailed view of the primary continuum along with the Fe K line features, which are the strongest in the spectrum. 
    We conservatively excluded all data above 9 keV from the analysis because their signal-to-noise ratio (S/N) is poor and they may be contaminated by background. 
    Subsequently, we extended the analysis to the soft X-rays, which allowed us to derive the best-fit parameters of the broadband spectrum in the observer frame $E=0.3-9$ keV band.
    We consider the errors on the individual parameters to be at 90\% confidence level. 
    We used all data from EPIC-pn and from the two MOS cameras.

    \subsection{Hard X-ray spectrum} 
    
    The spectrum is characterized by a remarkable absorption feature at $\sim 7.5$ keV, preceded by emission at $\sim 6.7$ keV.
    We started by introducing a baseline model consisting of a power law modified by Galactic absorption and an intercalibration constant between the different instruments.  
    The spectral fit returns a poor value of $\chi^2 = 466.48$ for 264 degrees of freedom (d.o.f.), and the power law has a photon index $\Gamma = 1.90 \pm 0.03$.
    
    \begin{figure}[t]
        \centering
        \includegraphics[width=\columnwidth]{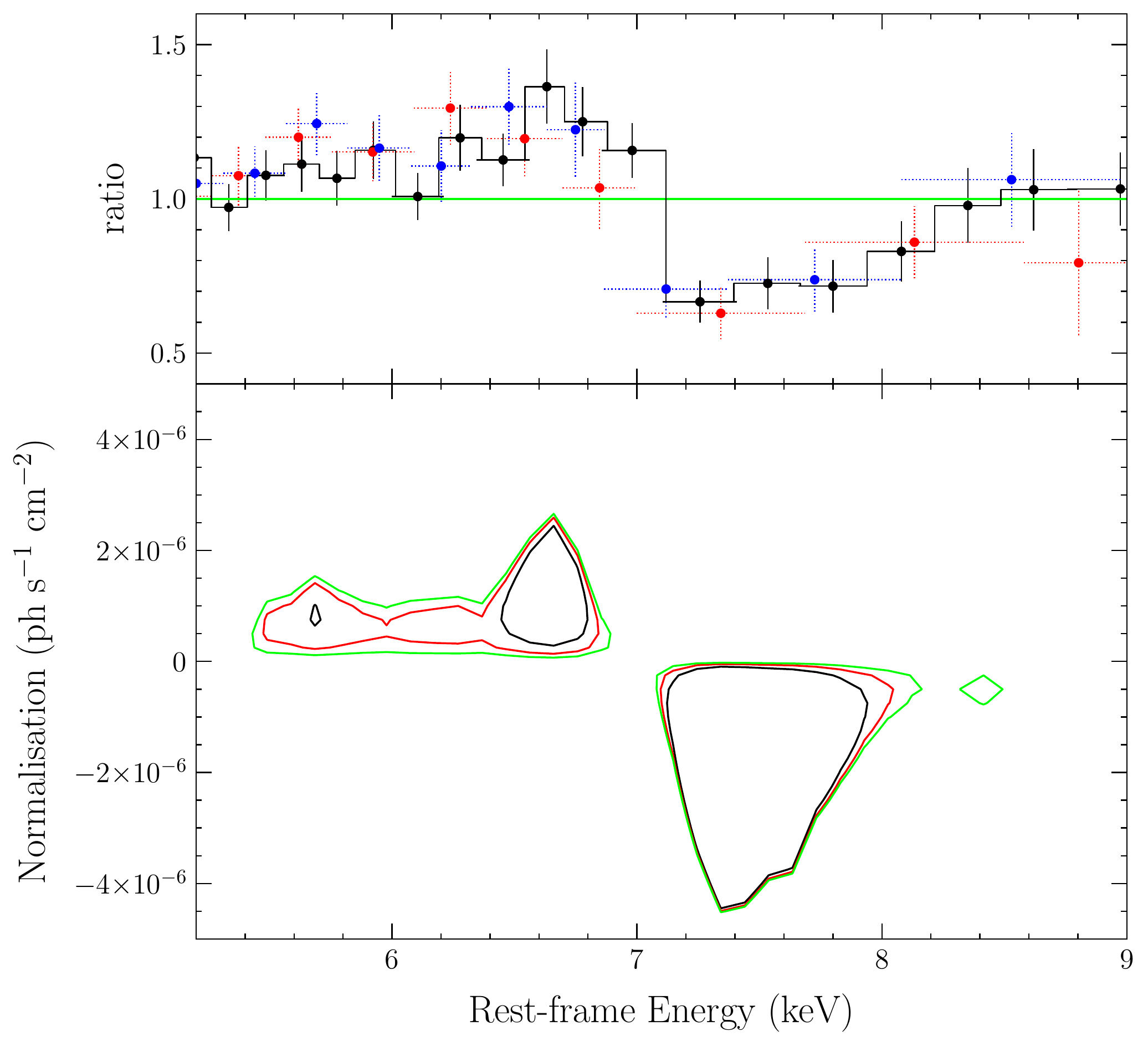}
        \caption{\emph{Top panel}: Ratio between the data points and the best-fit model.  Data from EPIC-pn, MOS1, and MOS2 are shown in black, red, and blue, respectively. The model consists of a single absorbed power law describing the continuum, zoomed in on the $E=4-9$ keV energy interval. Data are rebinned to a minimum S/N of $11\sigma$ and up to three counts only for plotting purposes. \emph{Bottom panel}: Contour plot describing the normalization vs. the rest-frame energy of a Gaussian line component in addition to the baseline model, consisting of a power law modified by Galactic absorption. Black, red, and green lines refer to the 68\%, 90\%, and 99\% confidence level, respectively. }
        \label{fig:zhard_and_cont}
    \end{figure}

    We show the ratio of the data points and the model in the top panel of Fig.\ \ref{fig:zhard_and_cont}. A strong absorption feature together with a weak preceding emission is clearly visible.
    To test this hypothesis, we added an unresolved Gaussian line to the baseline model, in which the line energy was free to vary between 5 and 9 keV, and the normalization was allowed both positive and negative values. We show in the bottom panel of Fig.\ \ref{fig:zhard_and_cont} the contour plot of these two parameters, indicating a clear absorption preceded by a weaker and less well resolved emission feature, in agreement with what we present in the top panel of the same figure.

    The fit improvement after the inclusion of an absorption line is substantial, with $\chi^2 = 308.18$ for 261 d.o.f.. The new value of the photon index is $\Gamma = 1.80 \pm 0.03$, and the absorption line has an energy $E = 7.51 \pm 0.07$ keV, a width $\sigma = 0.28_{-0.07}^{+0.08}$ keV, and an equivalent width $EW = -390 \pm 70$ eV.
    With respect to the baseline model, we find an improvement of $\Delta\chi^2 = 158.3$ for three model parameters. 
    The F-test indicates that the detection confidence level is $>9\sigma$. This confirms the line detection of \citet{kosec2020}, to which we refer for a detailed calculation of the Monte Carlo confidence level. Moreover, in the $E=7-8$ keV range, which is the region occupied by the absorption line, the background contribution is not substantial because the source counts are always more than $83\%$ of the total.
    
    We tested the significance of the line detection in each EPIC camera separately, and we show the results in Tab.\ \ref{tab:abstest}.
    We also find that the equivalent widths of the absorption line detected in each instrument are compatible within the errors. 
    These results further support the clear evidence that the absorption feature is a very robust detection. 
    
    As a further refinement of the hard X-ray spectral fit, we included a second Gaussian component to take the apparent emission feature at $E \sim 6.7$ keV into account. 
    We obtain best-fit values for the line energy $E = 6.67 \pm 0.08$ keV and equivalent width $EW = 90_{-30}^{+50}$ eV. The width of the line is unconstrained, and we fixed it to a value consistent with the energy resolution of the instruments of 100 eV. 
    The new parameters are $\Gamma = 1.82 \pm 0.03,$ and for the absorption line, the values are $E = 7.50 \pm 0.07$ keV, $\sigma = 0.3 \pm 0.1$ keV, and $EW = -410 \pm 80$ eV.

    \begin{table}[t]
    \centering
    \caption{Results of the test to confirm the absorption line profile in the EPIC-pn, MOS1, and MOS2 datasets. The fit improvement $\Delta\chi^2$ is always associated with $\Delta(\mathrm{d.o.f.}) = 3$.}
    \renewcommand{\arraystretch}{1.5}
    \begin{tabular}{c c c}
    \hline\hline
    Instrument & $\Delta\chi^2$  & $P_\mathrm{F}$  \\
    \hline
    pn & 72.58  & $>6\sigma$   \\
    MOS1 & 60.97  & $>5\sigma$   \\
    MOS2 & 39.14  & $>5\sigma$   \\
    pn + MOS1 + MOS2 & 158.3  & $>9\sigma$ \\
    \hline
    \end{tabular}
    \label{tab:abstest}
    \end{table}
    
    \subsection{Broadband X-ray spectrum}
    \label{broadband}

    The extension of the hard X-ray model to the broadband ($E=0.3-9$ keV observer frame) X-ray spectrum shows a strong soft excess that we modeled with a blackbody component. 
    The spectral fit returns a value of $\chi^2 = 763.02$ for 410 d.o.f.. We note that this result is affected by the complexity of the soft X-ray spectrum, which appears to display some possible contributions from both emission and absorption features below 1 keV \citep[see also][]{kosec2020}.
    The underlying continuum is described by a power law with $\Gamma = 1.89 \pm 0.02$ and a blackbody with $kT = 110.8 \pm 0.7$ eV.
    These values agree with the bulk of the PG sample of radio-quiet quasars (see e.g. \citealt{piconcelli2005} for an extensive review).
    We use this model in the next section to describe the ionizing continuum in the wind photoionization modeling. 
    
\section{Wind physical modeling}

\subsection{Model setup}
\label{modelsetup}
    To further investigate the Fe K emission and absorption profiles, we fit the hard X-ray spectrum with the novel WINE. It consists of a chain of calls to the X-ray photoionization code {\it XSTAR} \citep{kallman2001}, which calculates the radiative transfer inside the wind and allows us to obtain absorption and emission spectra. A preliminary version of the program has been presented in \citet[][]{luminari2018}, and a complete description of all its features will be provided in a forthcoming paper (A. Luminari et al., {\it in prep.}). 
    
    Input parameters for WINE are the incident spectrum $S_\mathrm{i}$ and $L_\mathrm{ion}$, the $1-1000$ Ry (1 Ry = 13.6 eV) integrated luminosity. The wind properties are described by the inner radius of the wind $r_0$, the ionization parameter $\xi_0$\footnote{$\xi_0$ is defined as $L_\mathrm{ion}/n_0 r_0^2$, where $n_0$ is the wind number density at $r_0$.}, the column density $N_\mathrm{H}$ , and the outflow velocity at $r_0$, referred to as $v_0$.
    
    \begin{figure}[t]
    \centering
    \includegraphics[width=\columnwidth]{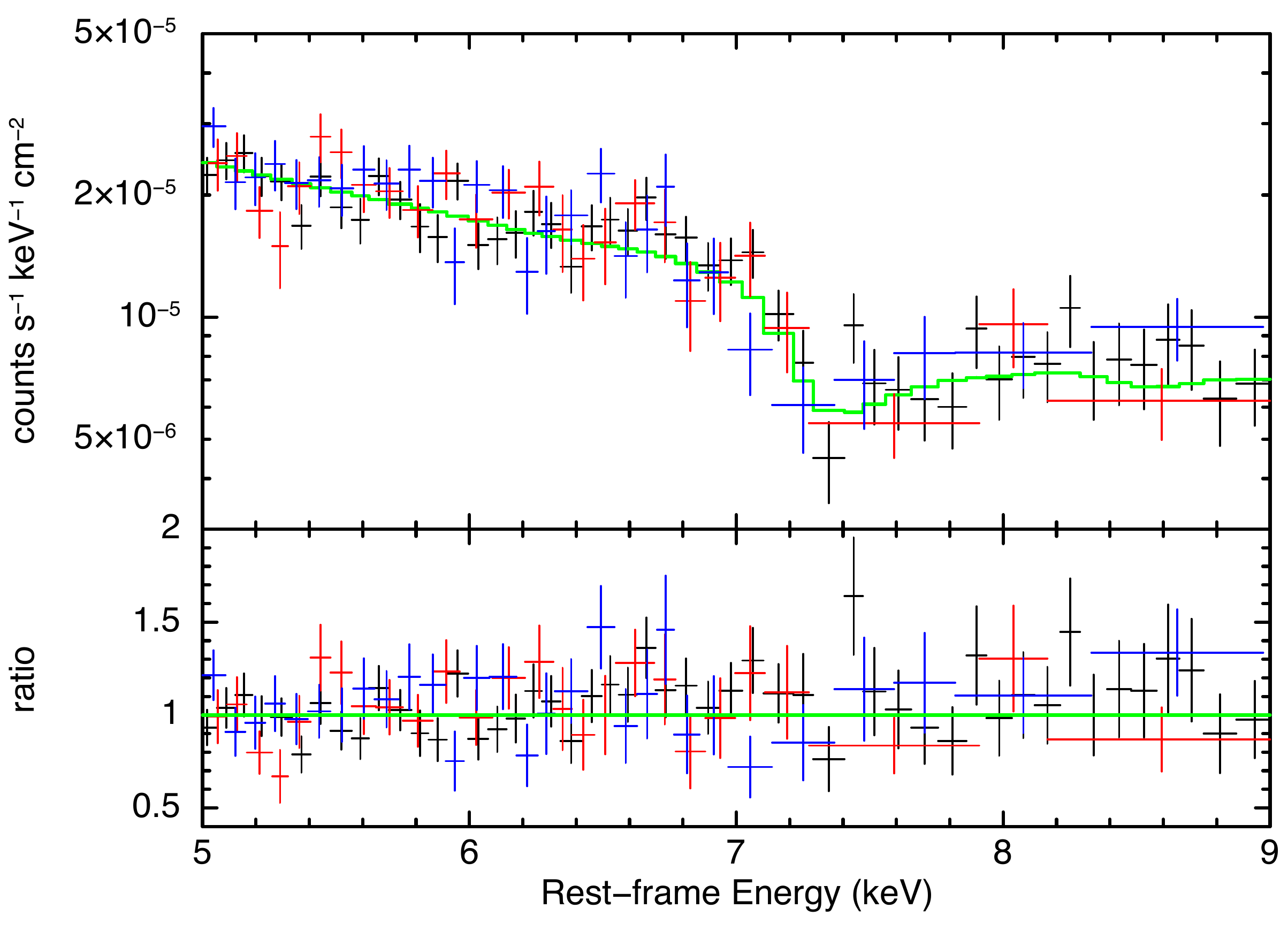}
    \caption{WINE spectral fit zoomed on the UFO absorption feature. Data from EPIC-pn, MOS1, and MOS2 are shown in black, red and blue, respectively. Residuals are presented in the bottom panel.}
    \label{fig:wine}
    \end{figure} 
    
    We defined the density and velocity profiles along the wind, $n(r)$ and $v(r)$, in the following way. As suggested by the absorption measure distribution \citep[AMD;][]{behar2009, tombesi2013} and by theoretical models within the magnetohydrodynamics (MHD) framework \citep[e.g.][]{fukumura2010}, we considered $n(r) = n_0\,(r_0/r)$.
    In the case of a momentum-conserving outflow, as expected for a fast wind lying at accretion disk scales, it is possible to link the radial dependence of $v(r)$ to $n(r)$ as $v(r) = v_0\,(r_0/r)^{1/2}$ (\citealp{faucher2012}, Appendix A; \citealp{king2015}). We note that this scaling of $v(r)$ is also  consistent with the case of a ballistic trajectory when the outflow, after it is launched with a given initial velocity, is subjected to the SMBH gravitational attraction during its expansion.
    In this way, we provide a self-consistent physical, kinematical, and geometrical picture of the wind with a limited number of parameters. This allows us to constrain all the parameters without the need to assume any value {\it \textup{a priori}}. 
    
    We calculated the absorption and emission spectra according to the following procedure. We divided the wind into equally spaced thin slabs with column $\delta N_\mathrm{H}$ or equivalently, radial thickness $\delta r = r_\mathrm{e} - r_\mathrm{i}$, where $r_\mathrm{i},\,r_\mathrm{e}$ are the starting and ending radii of the slab, respectively, and can be found by inverting the definition of $N_\mathrm{H}$. We set $\delta N_\mathrm{H}=10^{22}$ cm$^{-2}$ to obtain a good spectral energy resolution and an accurate scaling of $v,\, \xi$.
    We started the simulation from the innermost slab (i.e., for $r_\mathrm{i}=r_0$).
    Then, we used the output spectrum and luminosity, $S_\mathrm{out}$ and $L_\mathrm{out}$, as $S_\mathrm{i},\, L_\mathrm{i}$, for the following slab, and we iterated until the total column density $N_\mathrm{H}$ of the wind was reached.
    
    For each slab, we derived the ionization parameter at its inner edge as $\xi(r_\mathrm{i})=L_\mathrm{i}/n(r_\mathrm{i}) r_\mathrm{i}^2$. Importantly, we also included special relativistic effects, following the procedure outlined in \citet{Luminari20}. We used as outflowing velocity $v_\mathrm{out}=v(r')$, where $r'$ is the mass-averaged radial coordinate of the slab and is comprised of $r_\mathrm{i}$ and $r_\mathrm{e}$. 
    Finally, we defined the turbulent velocity $v_\mathrm{turb}$ (required by XSTAR) as the maximum between i) the velocity difference between adjacent shells and ii) an intrinsic turbulence, which we fixed to 10\% $v_\mathrm{out}$. We also tried different values for the intrinsic turbulence. However, because the wind absorption profile is governed by the wind velocity shear along the different shells, we did not note any appreciable difference.
    In this way, we obtained the special relativistic-corrected absorption spectrum of the wind. Moreover, we stored the atomic line emissivity files for each slab and used them to calculate the wind emission. In these files, the emissivity $e_\mathrm{j}$ (in units of luminosity per unit area, erg s$^{-1}$ cm$^{-3}$) of every atomic transition $j$ in the {\it XSTAR} atomic database is listed. We considered the wind-emitting region to have a conical shape, with the vertex centered on the SMBH and the same symmetry axis as the accretion disk (see Fig. 1 in \citealp{luminari2018}). We indicate the cone opening angle and the inclination of the line of sight (LOS) with respect to the symmetry axis with $\theta_\mathrm{out}$ and $i$, respectively.
    
    For each slab, we calculated the volume $A$ as a function of $r_\mathrm{i},r_\mathrm{e},\text{and } \theta_\mathrm{out}$. Assuming that the slab is emitting homogeneously, we calculated the total luminosity of the $m$ brightest lines in the hard X-ray energy interval as $E_\mathrm{j}=e_\mathrm{j}\times A$, where $m=10^6$ by default. 
    Then,  we assigned spatial coordinates to a large number $n$ ($n=10^4$) of points using a Monte Carlo method. We assigned to each point a line luminosity equal to $E_\mathrm{jn}=E_\mathrm{j}/n,$ and we calculated the luminosity projection and the blueshift along the LOS using special relativistic formulae. We randomly extracted the velocity of each point from a Gaussian distribution centered on $v(r')$ (i.e., the slab mass-averaged velocity) and with standard deviation $v_\mathrm{turb}$.
    We performed this calculation for all the $m$ brightest transitions, and we obtained the total emission from the point. Then, we repeated this operation for all the $n$ points to obtain the total slab emission spectrum.
    Finally, the total wind emission spectrum is given by the composition of all the slab spectra.
    \subsection{Results}
\begin{figure}[ht]
    \vspace{-0.1cm}
    \centering
    \includegraphics[width=\columnwidth]{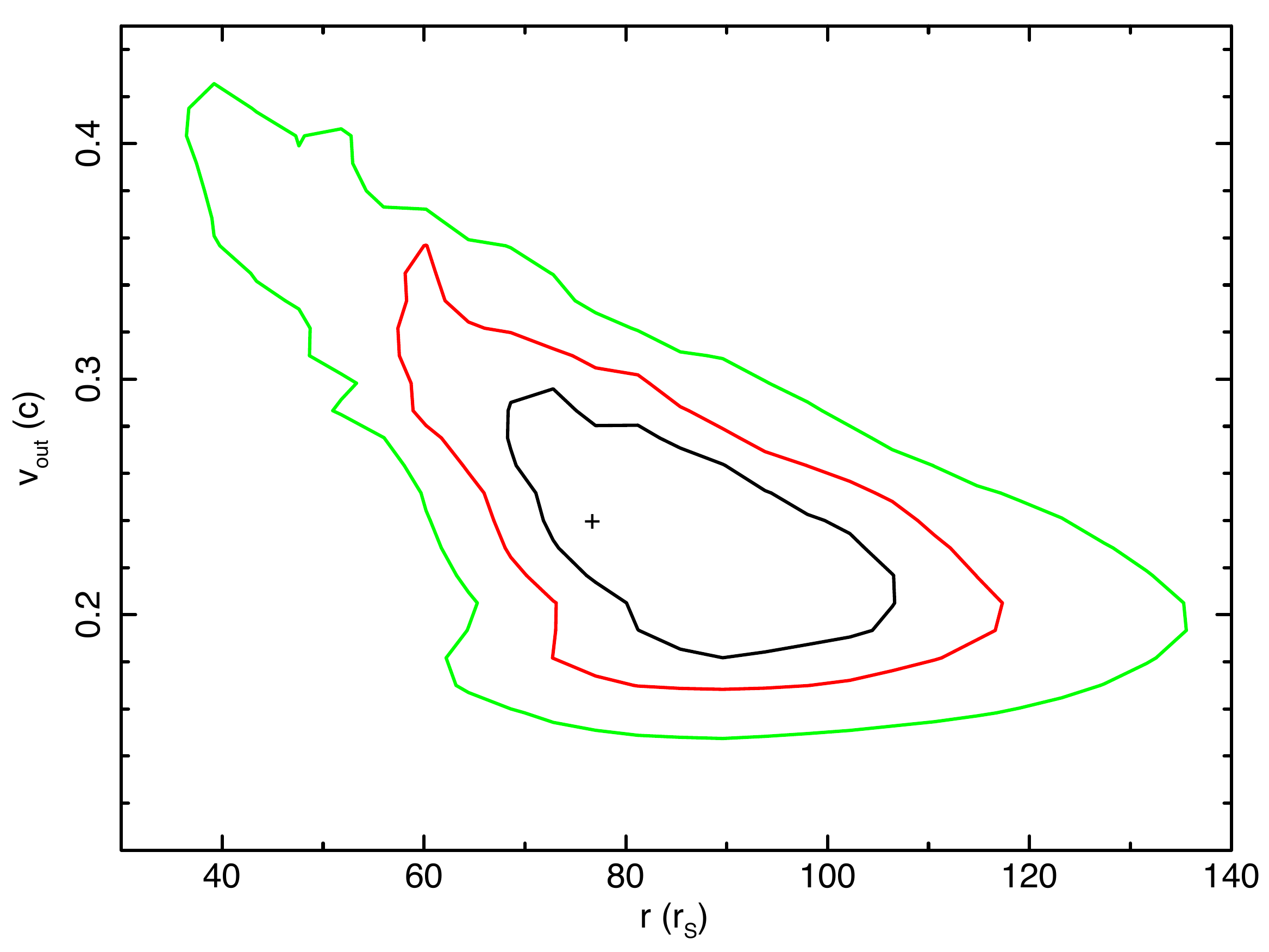}
    \includegraphics[width=\columnwidth]{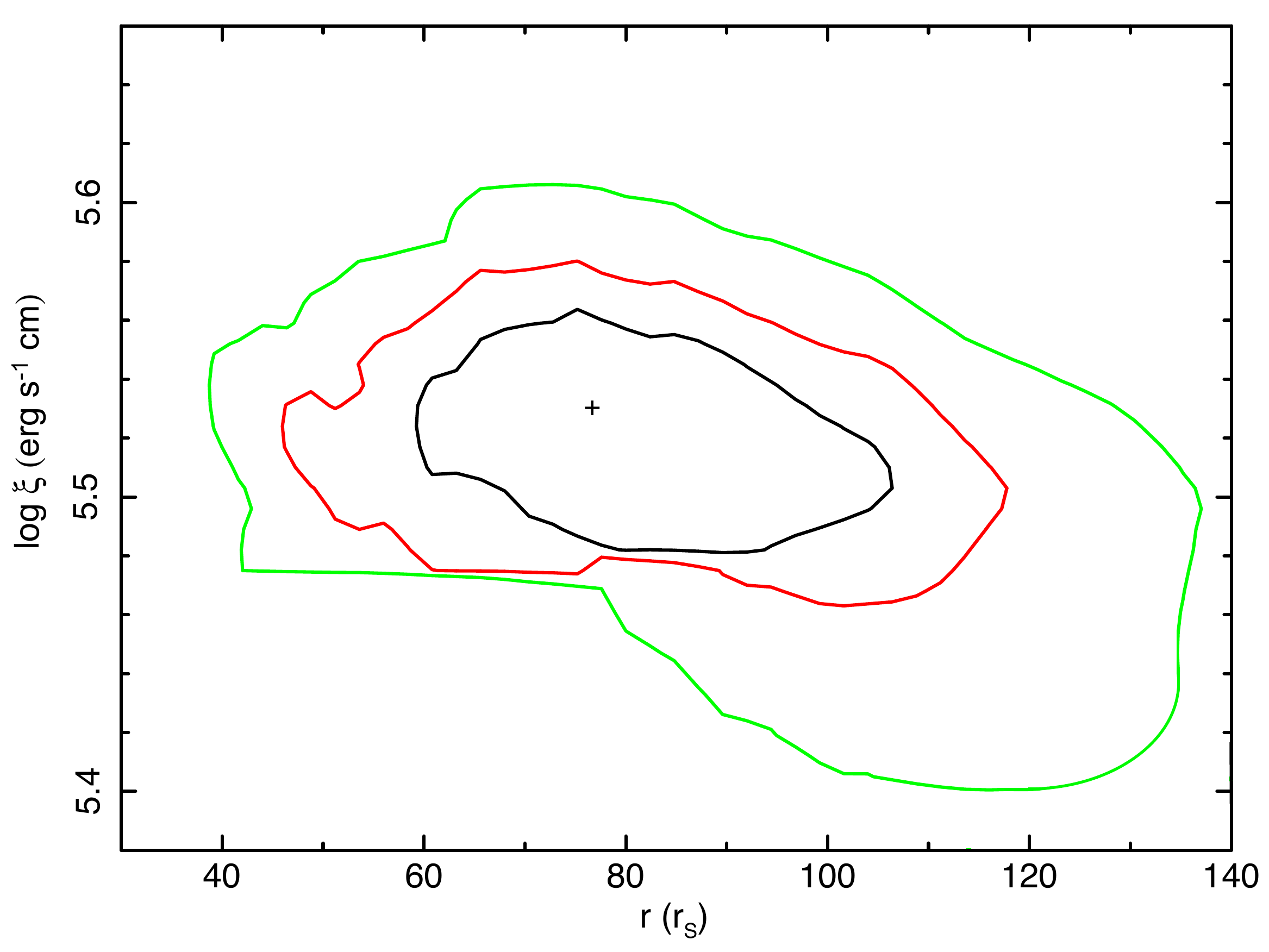}
    \includegraphics[width=\columnwidth]{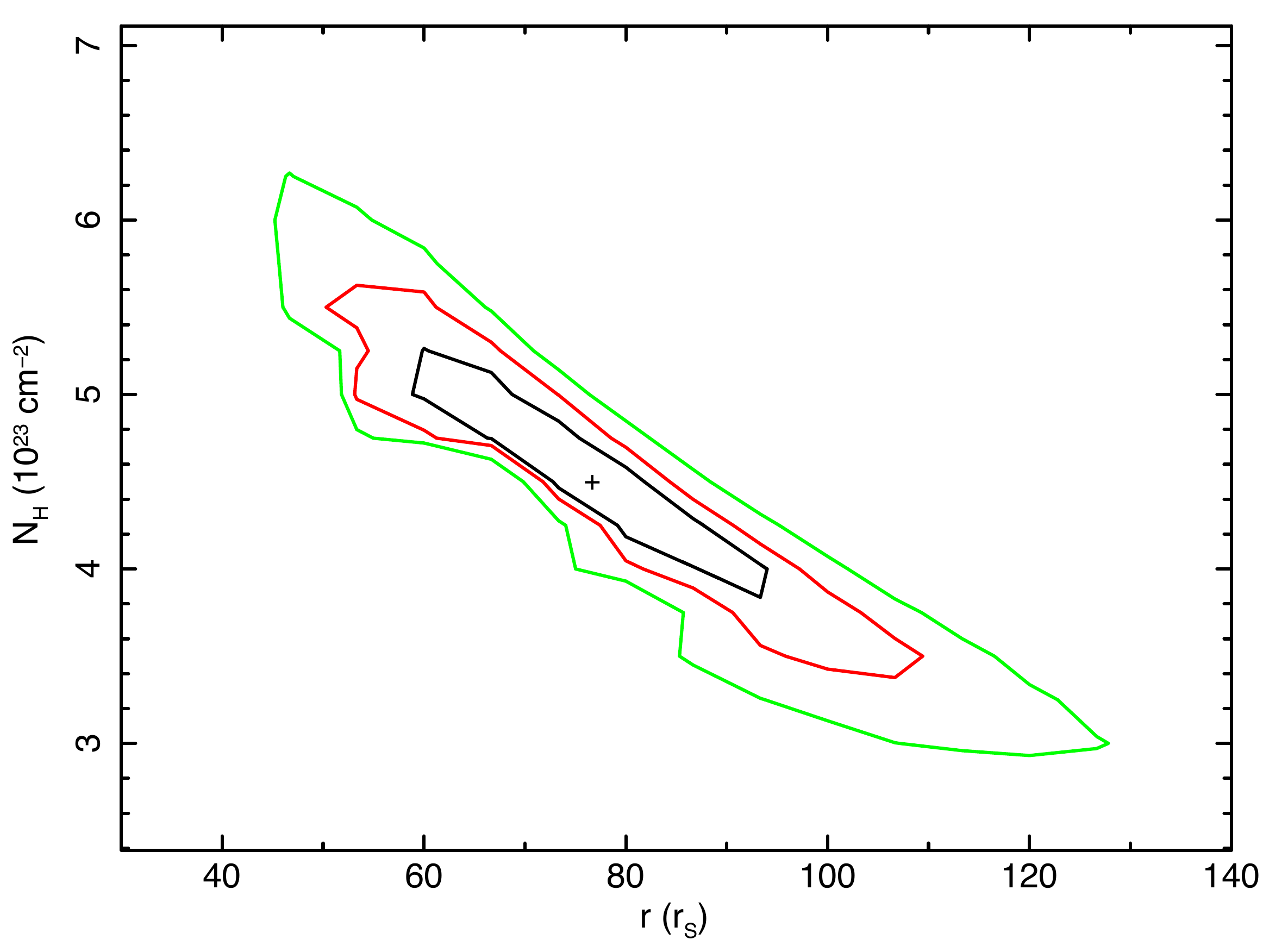}
    \caption{Contour plots of the launching radius vs. the outflow velocity (top panel), ionization parameter (mid panel) and wind column density (bottom panel). Black, red, and green lines correspond, respectively, to 68\%, 90\% and 99\% confidence level.}
    \label{fig:rv}
    \end{figure}

    We ran WINE using as incident spectrum and luminosity ($S_\mathrm{i} , L_\mathrm{ion}$, respectively) of the best-fit model of the broadband spectrum of PG 1448+273 (Sect. \ref{broadband}). Motivated by the results of the phenomenological fit, we explored the following intervals for the free parameters of the model:
    \begin{itemize}
        \item $r_0$: we spanned the interval $[10,\,140]\ r_\mathrm{S}$ with steps of 10 $r_\mathrm{S}$, where $r_\mathrm{S}= 2 GM / c^2$ is the Schwarzschild radius, and $G, M,\text{and } c$ are the gravitational constant, the SMBH mass, and the speed of light, respectively. This interval is consistent with the typical launching radius derived for UFOs (see, e.g., \citealp{tombesi2013,nardini2015}). We converted from $r_\mathrm{S}$ into centimeters using the black-hole mass listed in Tab. \ref{tab:infoPG}.
        \item $\xi_0$: we spanned the interval $\log{\xi_0} \in [4.0,\,6.0]$ with steps of 0.25. This high degree of ionization is consistent with a large fraction of Fe being in Fe XXV and Fe XXVI state \citep{kallman2004}, as suggested by the observed absorption and emission lines around $6-7$ keV.
        \item $N_\mathrm{H}$: we spanned the interval $[5\times10^{22},\,10^{24}]$ cm$^{-2}$ with steps of $5\times10^{22}$ cm$^{-2}$.
        \item $v_0$: we spanned the interval $[0.00,\,0.45]\,c$ with steps of 0.05$\,c$.
    \end{itemize}
    In the {\tt XSPEC} language the fitting model corresponds to the analytical expression
    \begin{equation}
        \mathit{constant*TBabs*( WINE_\mathrm{abs}*(zpow+zbb)+WINE_\mathrm{em}),}
    \end{equation}
    where $\mathit{WINE}_\mathrm{abs}$, $\mathit{WINE}_\mathrm{em}$ represent the wind absorption and emission modeled with WINE. 
    The continuum is described by a power law and a blackbody component (\textit{zpow} and \textit{zbb}, respectively), in agreement with the results of Sect.\ \ref{broadband}. \textit{TBabs} corresponds to Galactic absorption, and \textit{constant} represents an intercalibration constant between the different instruments. We concentrated on the hard X-ray spectrum to better model Fe K shell transitions, while leaving the photon index $\Gamma$ and the parameters of the black body component ($zbb$) fixed to the best values of the broadband fit. We fixed $i$, the inclination of the emission component, to $i=0\degree$ because it is not constrained by the fit to the data. In this way, we ensured that the LOS always intersected the wind, as expected from the detection of the absorption features imprinted by the outflowing gas. All wind parameters are linked between emission and absorption, with the only exception of $\theta_\mathrm{out}$, which belongs to the emission component alone. We list the best-fit values in Tab. \ref{WINE_values} along with the derived covering factor of the wind $C_\mathrm{f}$ and the mass-averaged wind velocity $v_\mathrm{avg}$. All errors are at 90\% confidence level unless otherwise stated. 
    The redshift of all the components was fixed to the value in Tab.\ \ref{tab:infoPG}.
    We show in Fig.\ \ref{fig:wine} a detail of the best fit with WINE, which shows that there are no residual structures. The model describes both the emission and absorption features with high accuracy. Contour plots between $r_0$ and $v_0, \log{\xi},\text{and } N_\mathrm{H}$ are displayed in Fig.\ \ref{fig:rv}.
    
    \begin{table}
    \caption{Best-fit values of the WINE model. All errors are at the 90\% confidence level except for $\theta_\mathrm{out}$ ($1 \sigma$). Notes: {\it (a)} Parameter linked with the corresponding parameter in the absorption component. {\it (b)} Parameter estimated with $1\sigma$ error. {\it (c)} Derived value.}
    \centering
    \renewcommand{\arraystretch}{1.3}
    \begin{tabular}{l c}
    
    \hline\hline
    Component & Value\\
    \hline
    \hline
    $wind_\mathrm{abs}$ \\
    $r_0$ ($r_\mathrm{S}$) & $77^{+31}_{-19}$ \\
    $\log{\xi}$ (erg s$^{-1}$ cm) & $5.53^{+0.04}_{-0.05}$ \\
    $N_\mathrm{H}$ (10\tu{23} cm$^{-2}$) & $4.5^{+0.8}_{-1.1}$ \\
    $v_0$ ($c$) & $0.24_{-0.06}^{+0.08}$ \\
    \hline
    $wind_\mathrm{em}$ \\
    $r_0$ ($r_\mathrm{S}$) & 77 $^a$\\
    $\log{\xi}$ (erg s$^{-1}$ cm) & 5.53 $^a$\\
    $N_\mathrm{H}$ (10\tu{23} cm$^{-2}$) & 4.5 $^a$\\
    $v_0$ ($c$) & 0.24 $^a$\\
    $\theta_\mathrm{out}$  ($\degree$) & $>72^b$ \\
    \hline
    $v_\mathrm{avg}$ ($c$) & $0.15^{+0.09}_{-0.08}$ $^c$ \\
    $C_\mathrm{f}$ & $>0.69$ $^c$ \\
    \hline\hline
    $\chi^2$/d.o.f. & 285.1/257 (=1.11) \\
    \end{tabular}
    \label{WINE_values}
    \end{table}
    
     We note that our results indicate a slightly higher ionized and more massive outflow than reported by \citet{kosec2020}, where $\log{\xi}=4.0 \pm 0.1$ erg s$^{-1}$ cm, and $N_\mathrm{H} = 2.8^{+1.2}_{-0.7} \times 10^{23}$ cm$^{-2}$;  the outflow velocity, $0.090 \pm 0.002\,c$, is instead consistent with our average $v_\mathrm{out}$ within the error bars.
The difference in $N_\mathrm{H}$ can be easily explained by taking the relativistic effects on the wind opacity into account, as done in WINE. Because the outflow moves with a relativistic $v_\mathrm{out}$, its apparent (i.e., observed) column density must be corrected for according to $v_\mathrm{out}$ to obtain the intrinsic $N_\mathrm{H}$ \citep{Luminari20}. Using the values reported in \citet{kosec2020}, we derive an intrinsic $N_\mathrm{H} = 3.5^{+1.5}_{-0.9} \times 10^{23}$ cm$^{-2}$, which agrees with the value reported in Tab.\ \ref{WINE_values}.
For the ionization degree, we observe an apparent positive correlation in \citet{kosec2020} between $\xi$ and the velocity dispersion $v_{\sigma}$ (see Fig.\ 5 in their paper). Their best-fit value for $v_{\sigma}$ is $2100$ km s$^{-1}$ $= 0.007 \,c$. In the WINE model, the broadening of the features is primarily governed by the velocity shear along the wind column (see the discussion in Sect.\ \ref{modelsetup}), and we can estimate the velocity dispersion as $v_{\sigma} \sim | v_0 - v_\mathrm{avg} | = 0.09\,c$ . Therefore we speculate that our $\xi$, $v_{\sigma}$ lie at the upper end of the observed correlation in \citet{kosec2020}. However, their table model only extends to maximum values of $\log{\xi}=4.8$ erg s$^{-1}$ cm, $v_{\sigma}=0.03\,c$, and therefore we cannot draw a firm conclusion on this point.

   \begin{figure}[t]
    \centering
    \includegraphics[width=\columnwidth]{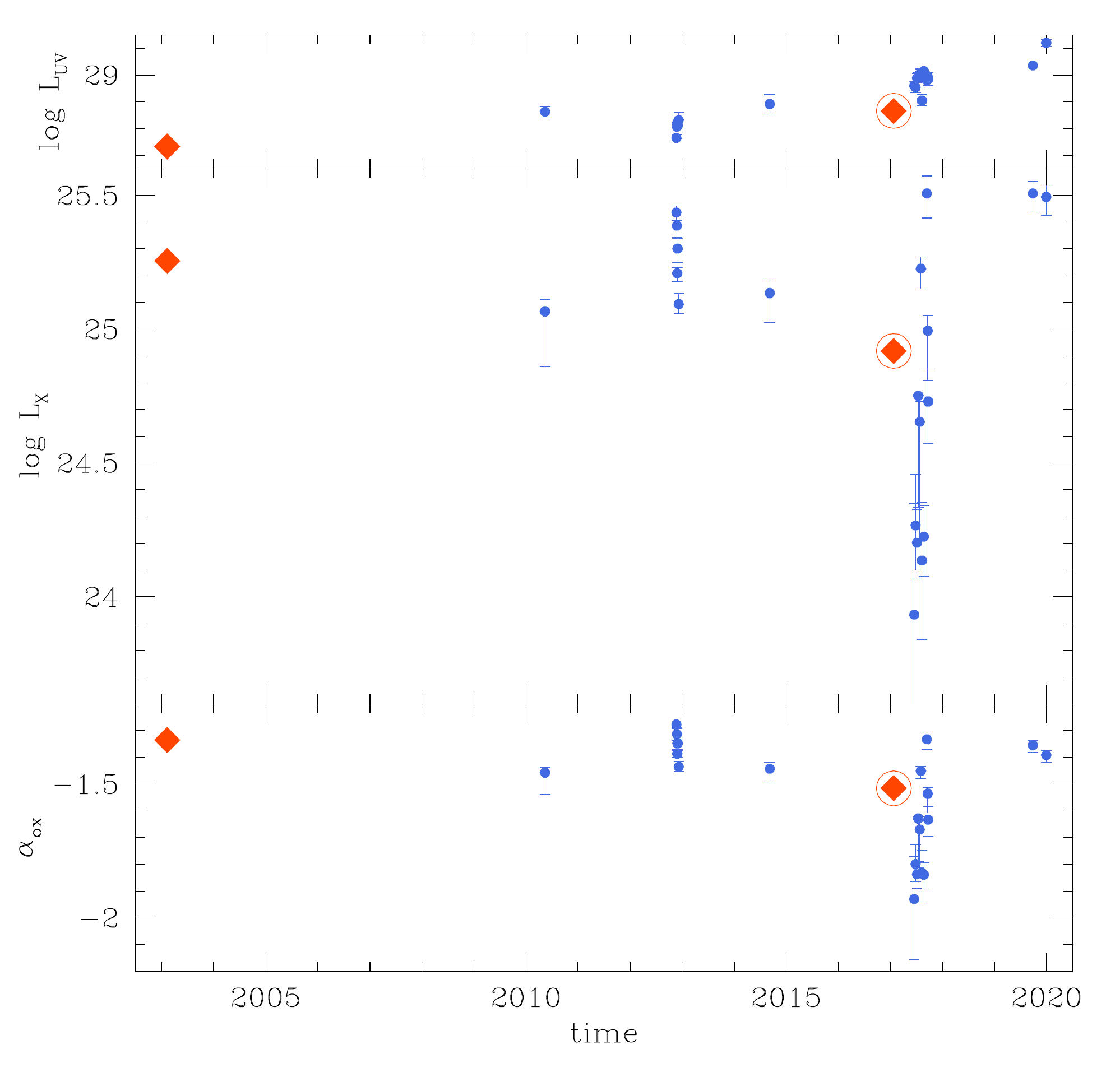}
    \caption{Light curves of the UV monochromatic luminosity at 2500\,\AA\ (top panel), the X-ray monochromatic luminosity at 2 keV (mid panel), the $\alpha_\mathrm{ox}$ parameter (bottom panel). Monochromatic luminosities are  in units of erg s$^{-1}$ Hz$^{-1}$. Red diamonds describe the \emph{XMM-Newton} observations. The diamond enclosed in a circle corresponds to the UFO detection. Blue points are related to the \emph{Swift} observations.}
    \label{fig:lightcurves}
     \end{figure} 
    
\section{Investigating the spectral and $\alpha_\mathrm{ox}$ variability}
\label{sec:aox}
Amplitude and spectral changes are hallmarks of AGNs. Variability occurs in all the wave bands and is witnessed from hours up to years and decades \citep[e.g.,][]{Green1993,Uttley2002,Ponti2012,Vagnetti2016,Paolillo2017,Middei2017}.
Ultraviolet and X-ray photons emerge from the innermost regions surrounding the central SMBH, that is, the accretion disk and the so-called hot corona \citep[e.g.,][]{Haardt1993}, and analyzing them allows us to probe the radiative balance between these two components.\\
\indent Simultaneous UV data are available from the OM pointings associated with the two \emph{XMM-Newton} observations. Three filters were used in the older 2003 observation (U, UVW1, and UVW2), and only one (UVW2) for the latest 2017 observation. For \emph{Swift}, six UVOT filters (V, B, U, UVW1, UVM2, and UVW2) are available for most of the observations, providing rich information on the slope of the UV continuum. This appears to be quite steep, with spectral index $\alpha$ (in the spectral region around 2500\,\AA) of between $-1.8$ and $-1$, which is far steeper than the typical value of $-0.57$ that we derive from the average spectral energy distribution (SED) by \citet{richards2006}. We ascribe this to a relevant contribution by the host galaxy emission. We determined the AGN luminosity at 2500\,\AA\ following \citet{vagnetti2013}, where we assumed a combination of an AGN spectrum, proportional to the average SED by \citet{richards2006}, plus a host galaxy spectrum with typical slope $-3$. This allowed us to estimate the fraction of host galaxy contribution as a function of the sole spectral index, which translates into an average host galaxy luminosity $\log L_\mathrm{G}=28.74$ (erg s$^{-1}$ Hz$^{-1}$), with a small dispersion $\sigma=0.06$ among the 20 \emph{Swift} epochs. This average value was then subtracted from the total UV luminosity to obtain the AGN luminosity $L_\mathrm{UV}$ at each epoch, which is shown in the top panel of Fig.\ \ref{fig:lightcurves}, which displays a moderate and nearly monotonic increase. 

Because of the AGN $L_\mathrm{UV}$ variation, the galaxy contribution evolves from 0.5 to 0.3.
Fig.~\ref{fig:lightcurves} also shows in the middle panel the light curve of the X-ray luminosity $L_\mathrm{X}$ at 2 keV, which is derived from the XRT observations.
In particular, XRT spectra were fit in the $E=0.3-10$ keV band with a power law absorbed for the Galactic column density as baseline model. Only when required by the data did we add a diskbb component to account for the soft excess. A sharp decrease in X-ray luminosity suddenly appears in 2017, and this is also marked by the behavior of $\alpha_\mathrm{ox}=0.38\log{(L_\mathrm{X}/L_\mathrm{UV})}$, shown in the bottom panel. The prominent X-ray variability observed in the middle panel of Fig.~\ref{fig:lightcurves} is found not to be driven by obscuration because the continuum spectra in the \emph{Swift} observations are similar. The variability is instead driven by an intrinsic decrease in X-ray flux.

\begin{figure}[t]
    \centering
    \includegraphics[width=\columnwidth]{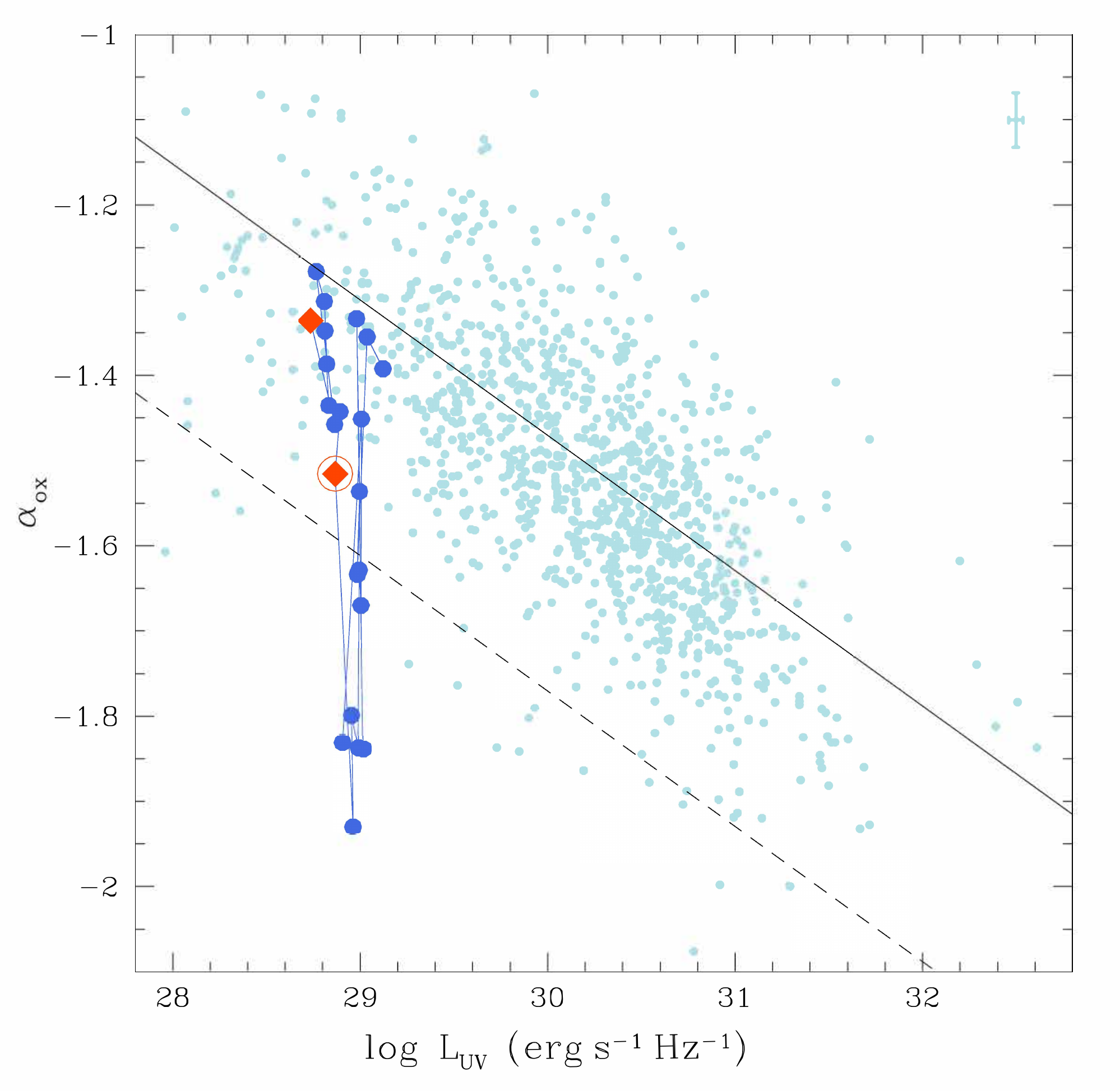}
    \caption{Track of PG 1448+273 in the $\log L_\mathrm{UV}-\alpha_\mathrm{ox}$ plane. The UV monochromatic luminosity is corrected for the host contribution, as described in the main text. Pale blue dots: Sample from \citet{chiaraluce2018}. Average error bars are showed in the top right corner. The solid black line represents $\log L_\mathrm{UV}-\alpha_\mathrm{ox}$ linear relation as derived by the same authors. Dashed black line marks the reference value for X-ray weakness, as discussed by \citet{pu2020}. The color scheme is the same as in Fig.\ \ref{fig:lightcurves}.}
    \label{fig:aloxluv}
    \end{figure}

\indent The $\alpha_\mathrm{ox}$ behavior can also be compared with the well-known $L_\mathrm{UV}-\alpha_\mathrm{ox}$ anticorrelation \citep[e.g.,][]{vignali2003,just2007,vagnetti2010}. Fig.\ \ref{fig:aloxluv} shows the track of PG 1448+273 in the $\log L_\mathrm{UV}-\alpha_\mathrm{ox}$ plane. The pale blue points in the background mark the region that is typically occupied by radio-quiet non-broad absorption line quasars, taken from \citet{chiaraluce2018}. Our quasar starts at $\alpha_\mathrm{ox}\approx -1.3$ decreases in 2017 to $\alpha_\mathrm{ox}\approx -1.9$ in the region of X-ray weak quasars \citep[e.g.,][]{brandt2000,miniutti2009,pu2020}, then it returns to standard values.
This trend is is reflected in the variation with time of the residual $\Delta\alpha_\mathrm{ox}(t)=\alpha_\mathrm{ox}(t)-\alpha_\mathrm{ox}(L_\mathrm{UV})$, where $\alpha_\mathrm{ox}(L_\mathrm{UV})$ is the expected $\alpha_\mathrm{ox}$ value according to the relation obtained by \citet{chiaraluce2018}. The residuals vary in the range $(0,-0.7),$ and many values are well below the limit of the so-called X-ray weak AGNs, which for example \citet{pu2020} set at $\Delta\alpha_\mathrm{ox}=-0.3$. It is worth mentioning that low values of $\alpha_\mathrm{ox}$ like this are observed about four months after the 2017 \emph{XMM-Newton} exposure characterized by the UFO.\\

\begin{figure}[t]
    \centering
    \includegraphics[width=\columnwidth]{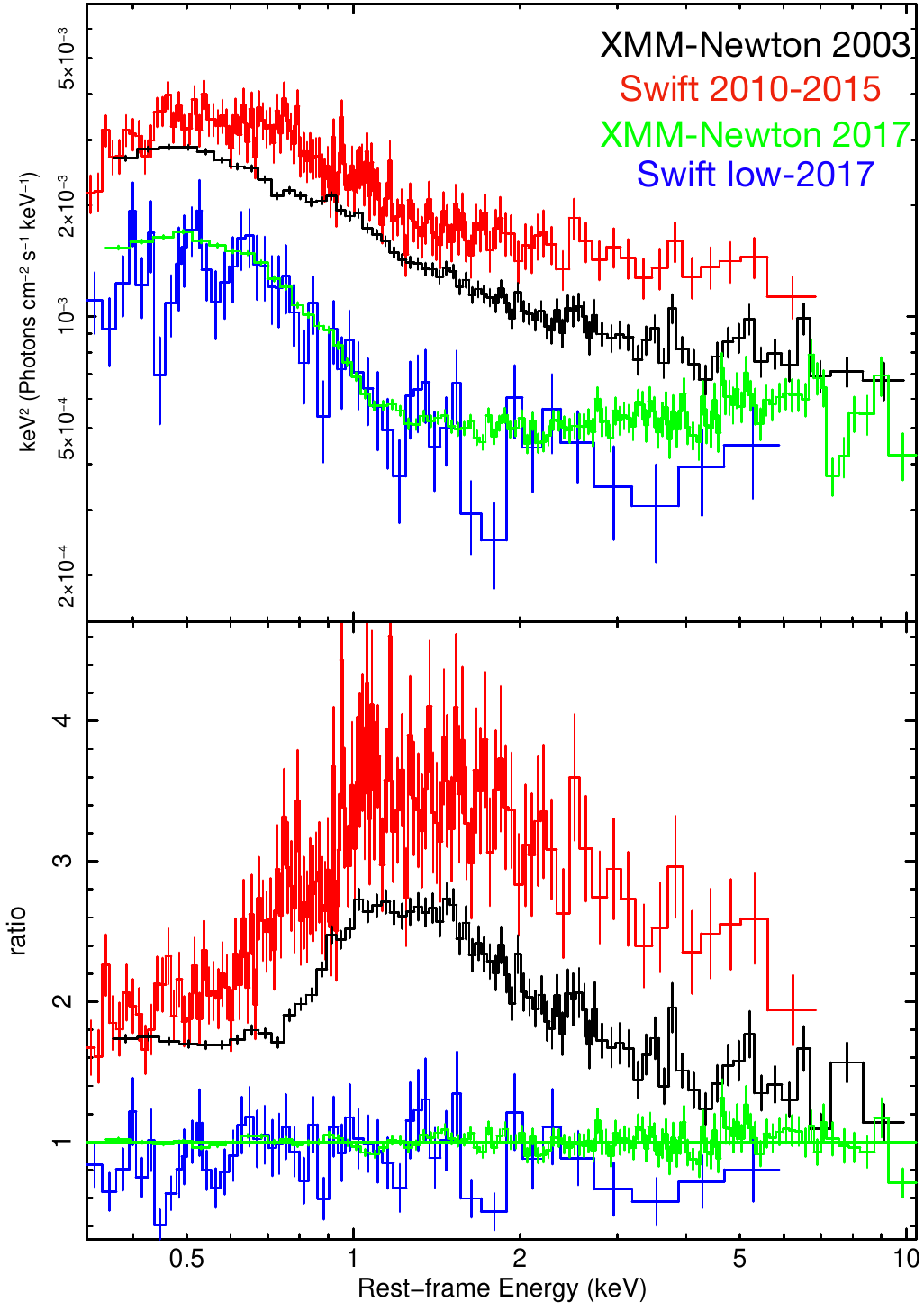}
    \caption{\emph{Top panel}: Unfolded \emph{XMM-Newton} spectra of 2003 (in black) and 2017 (in green) compared to the stacked \emph{Swift}/XRT spectra (in red and blue). In particular, the red spectrum is derived using all \emph{Swift}/XRT datasets between 2010 and 2015, while the blue spectrum is derived using low flux state ($\log{L_\mathrm{X}}<24.92$) \emph{Swift}/XRT spectra during 2017.
    \emph{Bottom panel}: Ratio of the data points of the \emph{XMM-Newton} observations, as well as the stacked \emph{Swift}/XRT observations, to a simple model describing the continuum of the 2017 \emph{XMM-Newton} observation. The color scheme is the same as in the top panel.}
    \label{pnxrt}
    \end{figure}

\indent These amplitude changes are also accompanied by variability in the source X-ray continuum shape. In top panel of Fig.~\ref{pnxrt} we show the unfolded \emph{XMM-Newton} spectra (black and green for 2003 and 2017 observations, respectively) and two \emph{Swift}/XRT stacked spectra. In particular, the red spectrum corresponds to the stacking of the \emph{Swift} data between 2010 and 2015 and the blue spectrum refers to the 2017 spectra in which the source was in a low state of luminosity $\log{L_\mathrm{X}}<24.92$, the value corresponding to the 2 keV monochromatic luminosity of the 2017 \emph{XMM-Newton} exposure. This panel shows that the power law in the hard ($> 2$ keV rest frame) X-rays agrees with the softer-when-brighter behavior, which is commonly observed in local Seyfert galaxies and distant quasars \citep[e.g.,][]{Sobolewska2009,Serafinelli2017}.
Spectral changes not only affect the primary continuum emission, but also act in the soft X-ray band. The bottom panel of the same figure shows the ratio of the spectra with respect to a simple phenomenological model accounting for the 2017 \emph{XMM-Newton} observations.

\section{Discussion} 

PG 1448+273 was observed twice by \emph{XMM-Newton}. The first observation dates back to February 2003, and the associated X-ray spectrum shows no peculiar features \citep[]{inoue2007}. In contrast, the observation of January 2017 reveals a clear UFO feature, as first noted by \citet{kosec2020}. Our independent study supports their finding, while examining in detail the physical properties of the wind with a novel modeling tool. We analyzed the UFO in PG 1448+273 by means of the WINE model, which allowed us to self-consistently derive the main properties of the UFO and to probe its geometry. 
We find that the outflowing material is highly ionized, $\log\xi = 5.53_{-0.05}^{+0.04}$ erg s$^{-1}$ cm, has a highcolumn density, $N_\mathrm{H} = 4.5_{-1.1}^{+0.8} \times 10^{23}$ cm$^{-2}$, is ejected with a maximum velocity $v_0 = 0.24^{+0.08}_{-0.06}\,c$, and attains an average outflow velocity $v_\mathrm{avg} = 0.152\,c$. WINE succeeds remarkably well in constraining the launching radius, which is $77^{+31}_{-19} \, r_\mathrm{S}$ away from the central SMBH, where $r_\mathrm{S}=2GM/c^2$ is the Schwarzschild radius. 
This constraint is of fundamental importance because it is often difficult to derive it in a stringent way, and usually only provide a rough estimate \citep[e.g.,][]{boissay2019} or upper and lower limits 
can be provided \citep[e.g.,][]{tombesi2012, gofford2015}.  
In addition, WINE allowed us also to derive a lower limit on both the opening angle of the wind ($\theta > 72\degree$) and the covering fraction ($C_\mathrm{f} > 0.69$).

Hereafter, we consider $1\sigma$ errors for the values of the wind energetics.
    Using the best-fit values of Tab.\ \ref{WINE_values}, we calculate the mass outflow rate of the UFO as
    \begin{equation}
          \Dot{M}_\mathrm{out}= 2\,\mu m_\mathrm{p} \int_0^{2 \pi } \int_0^{\theta_\mathrm{out}} \int_{r_0}^{r_1} n(r)\,v(r)\,r \sin(\theta)\ dr\, d\theta\, d\phi \
          \label{eq:mout}
   ,\end{equation}
    where $\mu, m_\mathrm{p}$ are the mean atomic mass per proton (which we set to 1.2, see \citealp{gofford2015}) and the proton mass, respectively. We derived $n_0$ by inverting the definition of $\xi_0,$ and we analytically calculated the ending radius of the wind $r_1$. We multiplied the integral by a factor 2 to take the contribution from both hemispheres into account. We obtain $\Dot{M}_\mathrm{out}=0.65^{+0.44}_{-0.33}\,M_\odot\,\mathrm{yr}^{-1} = 2.0^{+1.3}_{-1.0}\, \Dot{M}_\mathrm{acc}$ for a standard radiative efficiency $\eta=0.1$, as for the thin-disk model in \citealp{shakura1973}.
   We note that this value is a factor $\sim2$ higher than that derived using the formula in \citet{crenshaw2012} (see their Eq. 2), that is, $\Dot{M}^\mathrm{CK}_\mathrm{out}=4 \pi r_0 N_\mathrm{H} \mu m_\mathrm{p} C_\mathrm{f} v_0 = 0.25\pm 0.07 \,M_\odot$ yr$^{-1}$. This latter formula implicitly assumes a constant density outflow with a constant velocity, and it is consistent with those used in several UFO studies reported in the literature (e.g., \cite{tombesi2012,gofford2015,nardini2015,feruglio2015,dadina2018}). It corresponds to Eq.\ \ref{eq:mout} in the case of constant density, that is, $\alpha=0$ (where $n(r)=n_0 ( r_0/r )^{\alpha}$), and is characterized by a linear dependence on the wind parameters. In our work we instead set $\alpha=1$ (see Sect. \ref{modelsetup}), and as a result, the dependence on the parameters is nonlinear. This shows that the mean value and errors calculated with $\Dot{M}^\mathrm{CK}_\mathrm{out}$ may underestimate the real value of the wind energetics if its radial profile deviates from the $\alpha=0$ case.

   Moreover, we note that the main contribution to the uncertainties is given by $r_0$. The fractional uncertainty (i.e., the error interval divided by the mean value) for $r_0$ is the highest of the wind parameters in Tab.\ \ref{WINE_values}. 
   To assess the effect of the uncertainties of the fit parameters on $\Dot{M}_\mathrm{out}$, we calculated it by reducing the errors on each wind parameter in turn by a factor 10, while leaving all the others as in Tab.\ \ref{WINE_values}. While reducing the uncertainties on $\xi,\,N_\mathrm{H},\,v_0$ results in a decrease of less than 2\% in the errors of $\Dot{M}_\mathrm{out}$, reducing the error band on $r_0$ gives $\Dot{M}_\mathrm{out}=0.65_{-0.10}^{+0.14}\,M_\odot\,\mathrm{yr}^{-1}$, that is, a decrease of a factor of 4 and 2 for the lower and upper bound, respectively. This result shows the importance of an accurate determination of $r_0$. We also note that lower and upper bounds for $r_0$ are often derived in the literature with even order-of-magnitude differences between them (see e.g. \citealp{tombesi2013,gofford2015}), and this would result in even higher uncertainties for $\Dot{M}_\mathrm{out}$ .
   The increasing number of publicly available X-ray observations of UFOs provides the opportunity of constraining $r_0$ with higher accuracy through wind variability. Time-resolved spectroscopy for a set of X-ray observations of frequently monitored quasi-stellar objects that host a persistent UFO would enable improving the confidence intervals of the wind parameters and consequently, of the energetics. We plan to probe this approach in a future work. 
    
   We calculated the instantaneous kinetic power of the outflow using the special relativistic formula, as in \citet{saez2011},
   \begin{equation}
          \Dot{E}_\mathrm{out}= (\gamma -1)\cdot \Dot{M}_\mathrm{out} c^2 
   ,\end{equation}
    where $\gamma$ is the Lorentz factor, calculated using $v_\mathrm{avg}$, and we subtracted the energy at rest, $\Dot{M}_\mathrm{out} c^2$. We obtain $\Dot{E}_\mathrm{out}=4.4^{+4.4}_{-3.6} \times 10^{44}$ erg s$^{-1}$ = 24\% L$_\mathrm{bol}$ = 18\% L$_\mathrm{Edd}$. 
    Ultimately, we find that the outflow momentum rate is $\dot{p}_\mathrm{out} = 1.9 \times 10^{35} \text{ g cm s$^{-2}$} = 3.1\,\dot{p}_\mathrm{rad}$, where $\dot{p}_\mathrm{rad}=L_\mathrm{bol}/c$ is the radiation force.
    We now compared our results with those of \citet{kosec2020}. The authors derived the energetics of the wind (see their Eq.\ 2, 3, 4) using a formula for $\dot{M}_\mathrm{out}$ in the $\alpha=0$ case, similar to that in \citet{crenshaw2012}. Fixing the covering factor $C_\mathrm{V}=1$ and the filling factor $C_\mathrm{f}=1-\cos{(\theta_\mathrm{out})}$, with $\theta_\mathrm{out}$ as in Tab.\ \ref{WINE_values}, we find from their work $\dot{M}_\mathrm{out} = 0.48\,M_\odot$ yr$^{-1}$ and $\dot{E}_\mathrm{out} = 1.1 \times 10^{44}$ erg s$^{-1}$. As discussed before, we find $\dot{M}^\mathrm{CK}_\mathrm{out} = 0.25\,M_\odot$ yr$^{-1}$ and $\dot{E}_\mathrm{out} = 1.7 \times 10^{44}$ erg s$^{-1}$. The difference of about a factor of two in the results can be explained in terms of the different formula employed, as well as the different outflow velocities: $v_\mathrm{out} \simeq 0.09\,c$ for \citet{kosec2020} and $v_\mathrm{out} \simeq 0.15\,c$ in our study.

The instantaneous mechanical power of this UFO is large enough to drive a galactic-scale feedback, exerting a strong effect on the host galaxy, according to several simulations \citep[e.g.,][]{hopkins2010, gaspari2012}.
We also find that the outflow momentum rate exceeds  the radiation force by a factor $\sim 3$. 
This may possibly suggest that the role of the magnetic field is not negligible in accelerating the wind \citep[see, e.g.,][]{fukumura2015, kraemer2018, fukumura2018a, fukumura2018b}.
By using 20 \emph{Swift} (UVOT and XRT) observations together with the simultaneous OM data from \emph{XMM-Newton}, we also find that $\alpha_\mathrm{ox}$ undergoes large variations, with a maximum excursion of $\Delta\alpha_\mathrm{ox} =-0.7$ after the UFO is detected. A few months after the latest \emph{XMM-Newton} observation, $\alpha_\mathrm{ox}$  abruptly dropped toward very low values (down to $\alpha_\mathrm{ox} \approx -1.9$) that are typically associated with the X-ray weak quasars and then returned to the previous level (see Fig.\ \ref{fig:aloxluv}). The same behavior was observed for the X-ray luminosity (see Fig.\ \ref{fig:lightcurves}).
In principle, this steep decrease of $L_\mathrm{X}$ (and then $\alpha_\mathrm{ox}$) may be intimately connected with the outflow. Particularly, the UFO might subtract a significant fraction of the accreting material. The reduced accretion rate toward the innermost part of the accretion disk in which the corona originates would then lead to an X-ray weakness \citep[e.g.,][]{leighly2007,miniutti2009, martocchia2017,pu2020}.

An interplay between $L_\mathrm{X}$, $L_\mathrm{UV}$ and the outflows in the X-ray and UV band is observed in a growing number of sources \citep[see, e.g.,][]{nardini2019, zappacosta2020}. This indicates a coupling of these elements.
In particular, we note that $L_\mathrm{UV}$ moderately increases during the entire period between 2003 and 2019, even though with some scatter (Fig. \ref{fig:lightcurves}, top panel). Although the current observational sample is too small to draw any firm conclusion, we signal an interesting possibility that we plan to address with future complementary observations. If the long-term increase in $L_\mathrm{UV}$ is due to an increase in the mass-accretion rate onto the disk, then the onset of the UFO may have acted as an additional channel to remove angular momentum from the innermost part of the disk. If this is the case, the UFO represents a systemic component and can be used as a probe of the accretion-ejection physics.
Other X-ray and UV joint observations are required to shed further light on the accretion disk dynamics and its intimate relation with outflows.

        \begin{acknowledgements}
        We thank the anonymous referee for the useful comments, which helped us to improve our work.
           This work is based on observations obtained with \emph{XMM-Newton}, an ESA science mission with instruments and contributions directly funded by ESA Member States and NASA. Part of this work is based on archival data, software and online services provided by the Space Science Data Center - ASI.
           ML and AL thank MIUR (Ministero dell'Istruzione, dell'Universit\`a e della Ricerca) for the financial support to the Ph.D. programme in Astronomy, Astrophysics and Space Science. AL, FT and EP  acknowledge financial support under ASI/INAF contract 2017-14-H.0.
EP acknowledges support from PRIN MIUR project "Black Hole winds and the Baryon Life Cycle of Galaxies: the stone-guest at the galaxy evolution supper", contract \#2017PH3WAT.  RM acknowledges the financial support of INAF (Istituto Nazionale di Astrofisica), Osservatorio Astronomico di Roma, ASI (Agenzia Spaziale Italiana) under contract to INAF: ASI 2014-049-R.0 dedicated to SSDC.
        \end{acknowledgements}

\bibliographystyle{aa} 
\bibliography{39409corr.bib}
\begin{appendix}

\section{\emph{Swift} observations}

In Table \ref{app} we report the physical quantities derived from the \emph{Swift} observations.

\begin{table}[h]
\centering

\caption{\label{app} Journal of observations and derived physical quantities for the \emph{Swift} exposures.}
\renewcommand{\arraystretch}{1.5}
\begin{tabular}{c c c c c }
        
                \hline
        ObsID & Decimal year & $\log L_\mathrm{UV}$ (erg s$^{-1}$ Hz$^{-1}$) & $\log L_\mathrm{X}$ (erg s$^{-1}$ Hz$^{-1}$) & $\alpha_\mathrm{ox}$\\
\hline
\hline
  00037574001  &   2010.364 &   28.86 $\pm$0.02&    25.07 $_{- 0.05 }^{+ 0.21 }$ & $-$1.46 $_{- 0.02 }^{+ 0.08 }$\\
  00037574002  &   2012.888 &   28.77 $\pm$0.01&    25.44 $_{- 0.03 }^{+ 0.03 }$ & $-$1.28 $_{- 0.01 }^{+ 0.01 }$\\
  00037574003  &   2012.896 &   28.81 $\pm$0.05&    25.39 $_{- 0.02 }^{+ 0.04 }$ & $-$1.31 $_{- 0.02 }^{+ 0.02 }$\\
  00037574004  &   2012.904 &   28.82 $\pm$0.02&    25.21 $_{- 0.02 }^{+ 0.03 }$ & $-$1.39 $_{- 0.01 }^{+ 0.01 }$\\
  00037574005  &   2012.912 &   28.81 $\pm$0.02&    25.30  $_{- 0.04 }^{+ 0.05 }$& $-$1.35 $_{- 0.02 }^{+ 0.02 }$\\
  00037574006  &   2012.937 &   28.83 $\pm$0.03&    25.09 $_{- 0.04 }^{+ 0.04 }$ & $-$1.43 $_{- 0.02 }^{+ 0.02 }$\\
  00091340001  &   2014.682 &   28.89 $\pm$0.03&    25.14 $_{- 0.05 }^{+ 0.11 }$ & $-$1.44 $_{- 0.02 }^{+ 0.04 }$\\
  00037574007  &   2017.457 &   28.96 $\pm$0.02&    23.93 $_{- 0.41 }^{+ 0.59 }$ & $-$1.93 $_{- 0.16 }^{+ 0.23 }$\\
  00035079005  &   2017.485 &   28.95 $\pm$0.02&    24.27 $_{- 0.19 }^{+ 0.17 }$ & $-$1.80  $_{- 0.07 }^{+ 0.06 }$\\
  00035079006  &   2017.510 &   28.99 $\pm$0.02&    24.20  $_{- 0.12 }^{+ 0.14 }$& $-$1.84 $_{- 0.05 }^{+ 0.05 }$\\
  00035079007  &   2017.534 &   28.99 $\pm$0.02&    24.75 $_{- 0.01 }^{+ 0.42 }$ & $-$1.63 $_{- 0.01 }^{+ 0.16 }$\\
  00035079008  &   2017.567 &   29.00 $\pm$0.02&    24.65 $_{- 0.08 }^{+ 0.44 }$ & $-$1.67 $_{- 0.03 }^{+ 0.17 }$\\
  00035079009  &   2017.589 &   29.01 $\pm$0.01&    25.23 $_{- 0.05 }^{+ 0.08 }$ & $-$1.45 $_{- 0.02 }^{+ 0.03 }$\\
  00035079011  &   2017.605 &   28.91 $\pm$0.02&    24.14 $_{- 0.22 }^{+ 0.29 }$ & $-$1.83 $_{- 0.08 }^{+ 0.11 }$\\
  00035079012  &   2017.644 &   29.01 $\pm$0.01&    24.23 $_{- 0.12 }^{+ 0.15 }$ & $-$1.84 $_{- 0.04 }^{+ 0.06 }$\\
  00035079013  &   2017.701  &   28.98 $\pm$0.03&    25.51 $_{- 0.07 }^{+ 0.09 }$ & $-$1.33 $_{- 0.03 }^{+ 0.04 }$\\
  00035079014  &   2017.720 &   29.00 $\pm$0.02&    24.99 $_{- 0.06 }^{+ 0.19 }$ & $-$1.54 $_{- 0.02 }^{+ 0.07 }$\\
  00093092002  &   2017.726 &   28.98 $\pm$0.02&    24.73 $_{- 0.12 }^{+ 0.16 }$ & $-$1.63 $_{- 0.05 }^{+ 0.06 }$\\
  00095087001  &   2019.737 &   29.04 $\pm$0.01&    25.51 $_{- 0.05 }^{+ 0.07 }$ & $-$1.35 $_{- 0.02 }^{+ 0.03 }$\\
  00095087002  &   2019.994 &   29.12 $\pm$0.01&    25.49 $_{- 0.04} ^{+ 0.07 }$ & $-$1.39 $_{- 0.02 }^{+ 0.03 }$\\
  \hline
\end{tabular}

\end{table}

\end{appendix}

\end{document}